\documentclass[12pt]{article}
\usepackage{epsfig}

\usepackage{amssymb}
\usepackage{amsmath}
\usepackage{amsfonts}

\usepackage{amsthm}

\def\be{\begin{equation}}
\def\ee{\end{equation}}
\def\ba{\begin{array}{c}}
\def\ea{\end{array}}

\def\ben{$$}
\def\een{$$}
\newcommand{\bbr}{\br\!\br}

\newcommand{\kt}{\rangle}
\newcommand{\br}{\langle}
\begin{document}

%TITLEPAGE .

\vspace{.35cm}

 \begin{center}{\Large \bf

Non-Hermitian star-shaped quantum graphs\footnote{Material presented
during the AAMP 10 conference in Prague\\ (Villa Lanna, June 4 - 7,
2012,
http://gemma.ujf.cas.cz/\~{}znojil/conf/micromeetingdeset.html)}

  }\end{center}

\vspace{10mm}

 \begin{center}

 {\bf Miloslav Znojil}

 \vspace{3mm}
Nuclear Physics Institute ASCR,

250 68 \v{R}e\v{z}, Czech Republic

{e-mail: znojil@ujf.cas.cz}

\vspace{3mm}

%\today, vegu.tex

\end{center}

\vspace{5mm}

%\newpage

\section*{Abstract}

A compact review is given, and a few new numerical results are added
to the recent studies of the $q-$pointed one-dimensional star-shaped
quantum graphs. These graphs are assumed endowed with certain
specific, manifestly non-Hermitian point interactions, localized
either in the inner vertex or in all of the outer vertices and
carrying, in the latter case, an interesting zero-inflow
interpretation.

\newpage

%\section{Introduction \label{I} }

\section{The concept of quantum graphs}

The scope of the traditional quantum theory is currently being
broadened to cover not only the 1D motion of (quasi)particles, say,
along a single thin wire  but also along the more complicated
one-dimensional graph-shaped structures $\mathbb{G}$ which are, by
definition, composed of some $N_W$ (finite or semi-infinite) wedges
which may -- though need not -- be connected at some of the elements
of the given set of $N_V$ vertices. Under the influence of the
contemporary solid-state technologies these structures may be
realized as ultrathin 1D-like waveguides with nontrivial
topological, geometrical and/or physical properties. Nowadays, there
exists an extensive literature on the subject (cf., e.g.,
\cite{Exner}).

From the formal point of view, one of the key features of these
structures is the necessity of their description using the formalism
of Quantum Mechanics. This necessity opens, naturally, a number of
challenging questions which concern both the kinematics (for this
reason, in a typical quantum graph model one assumes a free motion
along the interior of the edges) and the dynamics (most often, one
uses just point interactions which are localized strictly at the
vertices).

\section{The concept of crypto-Hermiticity}

One of the most interesting mathematical questions addressed in the
quantum-graph context concerns the proofs or the sufficient
conditions of the self-adjoint nature of the corresponding
Hamiltonian $H=H^{(\mathbb{G})}$ in a pre-selected, sufficiently
``friendly'' Hilbert space ${\cal H}^{(F)}$ of states. Recently, a
new approach to this type of problems (carrying a nickname of ${\cal
PT}-$symmetric Quantum Mechanics) has been advocated and made
popular by Carl Bender with multiple coauthors (cf. review
\cite{Carl}).

The mathematical essence of their recommendation may be seen in the
replacement of the {\em a priori} selected friendly or ``first''
Hilbert space ${\cal H}^{(F)}$ by another, {\em ad hoc} and
Hamiltonian-dependent ``second'' Hilbert space ${\cal H}^{(S)}$
which only differs from ${\cal H}^{(F)}$ by a redefinition of the
inner product defined in terms of an operator $\Theta$ called
``Hilbert space metric'',
 \be
 \br \phi|\psi\kt^{(S)} := \br \phi|\Theta| \psi\kt^{(F)}\,\equiv\,
 \bbr \phi | \psi\kt^{(F)}\,
 \ee
(we use the Dirac-like notation conventions as introduced in the
compact reviews \cite{SIGMA,jakub}).

As an immediate consequence one reveals, in general, that one may
define a generic quantum system by a doublet of operators (i.e., by
the Hamiltonian $H$ and metric $\Theta$) which only have to satisfy
the following generalized version of the standard and necessary
Hermiticity requirement
 \be
  H^\dagger\,\Theta=\Theta\,H\,,
  \ \ \ \ 0 < \Theta=\Theta^\dagger < \infty\,
  \label{dieudo}
   \ee
which may be called a ``disguised'' Hermiticity or
crypto-Hermiticity \cite{Smilga}. The introduction of this concept
may be attributed, historically, to Dieudonn\'e \cite{Dieudonne}
and, in the context of (nuclear) physics, to Scholtz et al
\cite{Geyer}.

\section{The concept of crypto-Hermitian quantum graphs}

Even for the most trivial, $N_W=1$ quantum graphs with
$\mathbb{G}=\mathbb{R}$ it is by far not easy to satisfy the
selfconsistency condition (\ref{dieudo}). In essence, two technical
problems are being encountered. Firstly, we must guarantee that the
energy spectrum of our Hamiltonian is real (this is necessary for
probabilistic interpretation/tractability but, generically, the
proof is difficult -- see \cite{DDT} for an illustrative example).

Secondly, the compatibility (\ref{dieudo}) between $H$ and $\Theta$
(studied, most often, as a brute-force construction of
$\Theta=\Theta(H)$) is usually not an easy task, either (for the
above-mentioned example, just various tedious approximative
constructions may be found in the literature  -- cf.
\cite{cubic}).

This being said, one of the most efficient ways of addressing these
technical challenges has been described in Ref.~\cite{fund} and
generalized, to nontrivial graphs, in its sequel~\cite{fundgra}. The
guiding idea lied in the discretization of the graphs (see also
Ref.~\cite{katka} for a deeper analysis and discussion of this
technique). Its use opened also the successful transition to the
description of some topologically nontrivial graphs in
Refs.~\cite{anomalous}.

Unfortunately, for the topologically more complicated
crypto-Hermitian graphs, the advantages of the discretization
technique (which, basically, converted the underlying
Schr\"{o}dinger equation into a finite-dimensional matrix problem)
proved more or less lost due to the growth of the difficulties
caused by the necessity of finding a suitable metric
$\Theta=\Theta(H)$.

In this sense, the discrete elementary solvable examples as
described in Refs.~\cite{fund} and \cite{fundgra} appeared
exceptional and exceptionally friendly. For this reason we returned,
recently, to the study of the continuous quantum graphs again. In
order to keep them tractable non-numerically, we restricted our
attention to the mere topologically trivial, star-shaped graphs in
Ref.~\cite{archiv}. Here we intend to recall one of these models and
add a few comments, yielding the conclusion that the study of the
${\cal PT}-$symmetric quantum graphs may be made sufficiently
friendly not only via the simplification of kinematics (i.e., by the
discretization of the wedges) but also via the alternative extreme
simplification of dynamics. Not realized  by the selection of the
sufficiently elementary form of the graph itself but rather by the
use of the symmetry-rich point interactions at the vertices.

\section{The simplifications of kinematics}

During the study of the broad variety of the graphs-related open
questions one often selects the models where the underlying
mathematics is  simplified, to its very core, by the replacement of
the real-world problem in 3D by its idealized representation by a
piecewise linear 1D graph, on the wedges of which the Hamiltonian is
represented by the Laplacean, $H \sim -\triangle$. In such a
scenario one may feel inspired by the simplest systems which live on
the single straight line which may be finite (then, the system is,
in effect, a square well possessing just bound states) or infinite
(then, one usually speaks about the one-dimensional problem of
scattering).

\begin{figure}[h]                     %instead of \begin{figure}[t]
\begin{center}                         %instead of \begin{center}
\epsfig{file=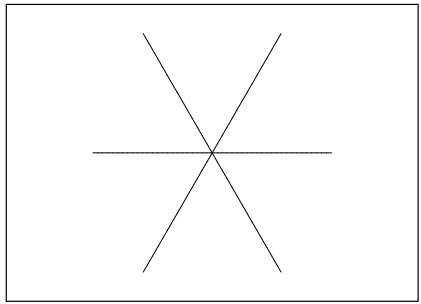,angle=270,width=0.4\textwidth}
\end{center}                         %instead of \end{center}
\vspace{-2mm} \caption{A typical six-pointed star graph.
 \label{fixu}}
\end{figure}

In the most natural direction of a generalization of the theory one
then glues a $q-$plet of half-lines in a single vertex and arrives
at the star-shaped family of graphs $\mathbb{G}^{(q)}$ (see
Fig.~\ref{fixu} where $q=6$ and where all of the wedge lengths and
angular distances of neighbors are equal).

%loweral.txt
%\newpage
%********** Figure 0 zde

Naturally, the main principles of Quantum Mechanics remain
unchanged. In particular, the bound states living on a given star
graph $\mathbb{G}^{(q)}$ will be represented by the quadratically
integrable wave functions $\psi^{}_n(x) \in L^2(\mathbb{G})$,
 \be
 \int_{\mathbb{G}} dx\,
 \psi^{*}_n(x)
 \,\psi^{}_n(x)\ \ < \ \ \infty\,.
 \label{jednac}
 \ee
Whenever the corresponding friendly and trivial metric
$\Theta^{(F)}=I$ is replaced by its nontrivial version defining the
correct physical Hilbert space ${\cal H}^{(S)}$, the ``naive''
Eq.~(\ref{jednac}) must be replaced by the double integral which
defines, by the way, also the correct physical orthonormalization of
the bound states,
 \be
 \int_{\mathbb{G}}  dx\,\int_{\mathbb{G}}
 dx'\,\psi_m^*(x)\,\Theta_{(x,x')}\,\psi_n(x')
 =\delta_{mn}\,
 \label{laots}
 \ee
where $\delta_{mn}$ is Kronecker symbol. For the Hamiltonians which
are Hermitian in ${\cal H}^{(F)}$ the use of the so called Dirac's
metric $\Theta_{(x,x')}=\delta(x-x')$ leads  to the degenerate form
(\ref{jednac}) of Eq.~(\ref{laots}) of course.

In the most natural move towards a more realistic scenario one often
tries to think about a nonzero thickness of the wedges. Of course,
with the necessary transition to the partial-differential
Schr\"{o}dinger equations this would make the eigenvalue problem
highly nontrivial even for the straight-line 1D graph. In
Ref.~\cite{fund}, an alternative idea of the smearing of the 1D
wedges has been proposed, therefore. Its realization relied heavily
upon the discretization of the 1D graph which reduced
Eq.~(\ref{dieudo}) to the mere $N$ by $N$ matrix problem. The key
idea lied in the smearing of the wedges caused by the use of the
{\em smearing-providing metric} $\Theta$. Illustrative examples were
constructed in the form of the tridiagonal (or block-tridiagonal or
block-pentadiagonal, etc) matrices $\Theta$ (see also
Ref.~\cite{smeared} for a more detailed account of this
metric-mediated smearing idea as well as of its possible
phenomenological interpretation and consequences).

\section{The simplifications of dynamics}

The discretization of the star-shaped graphs rendered it possible to
construct the smearing-metric matrices $\Theta$ for various models
of dynamics. For illustration, we may recall the one-parametric
toy-model discrete-graph Hamiltonian of Ref.~\cite{fund} with $N=4$,
 \be
 H^{(4)}(\lambda)=
 \left [\begin {array}{cccc} 2&-1&0&0\\{}-1&2&-1-\lambda&0
\\{}0&-1+\lambda&2&-1\\{}0&0&-1&2
\end {array}\right ]\,,
\label{cetyrki}
 \ee
which has been assigned there the four-parametric set of metrics
 \be
 \Theta^{(4)}(\lambda)
 =\Theta^{(4)}_{[\alpha_1,\alpha_2,\alpha_3,\alpha_4]}(\lambda)=
 \alpha_1\,M_1+\alpha_2\,M_2+\alpha_3\,M_3+\alpha_4\,M_4\,
 \label{super4}
  \ee
where
 \ben
 M_1=
 \left [\begin {array}{cccc} 1-\lambda&0&0&0
 \\{}
 0&1-\lambda&0&0\\{} 0&0&1+\lambda&0\\{} 0&0&0&1+\lambda\end {array}\right ]\,,\ \ \ \ \ \
 M_2=
 \left [\begin {array}{cccc}
  0&1-\lambda&0&0\\{}
 1-\lambda&0&1-\lambda^2&0
 \\{}
 0&1-\lambda^2&0&1+\lambda\\{} 0&0&1+\lambda&0\end {array}\right ]\,,
 \een
 \be
 M_3=
 \left [\begin {array}{cccc}0&0&1&0\\{}0&1-\lambda&0&1\\{}
  1&0&1+\lambda&0
 \\{}
 0&1&0&0\end {array}\right ]\,,\ \ \ \ \ \
 M_4=
 \left [\begin {array}{cccc}
 0&0&0&1\\{} 0&0&1&0\\{}
 0&1&0&0\\{}
  1&0&0&0
  \end {array}\right ]\,.
  \label{sihi}
 \ee
Obviously, in the light of the well known spectral-expansion formula
\cite{kappa} the general $N$ by $N$ matrix of the metric contains
$N$ free real parameters so that due to the independence of
components (\ref{sihi}) the solution (\ref{super4}) of
Eq.~(\ref{dieudo}) is exhaustive whenever all the four $\alpha_j$s
are positive and whenever $|\lambda|<1$. Moreover, the first
component $M_1$ is diagonal (so that we may conclude that the energy
spectrum remains real). At the same time, merely the last component
{\em cannot} be treated as a smearing-mediating band matrix.

Analogous (though, naturally, not so easily displayed) results were
obtained, in \cite{fundgra}, for the nontrivial {\em discrete} star
graphs with $q>2$. In place of the diagonal, tridiagonal,
pentadiagonal (etc) matrices (cf Eq.~(\ref{sihi})) we merely had to
construct there the block-diagonal, block-tridiagonal,
block-pentadiagonal (etc) matrices, respectively.

\section{An illustrative, exactly solvable star-shaped 1D quantum graph}

\begin{figure}[h]                     %instead of \begin{figure}[t]
\begin{center}                         %instead of \begin{center}
\epsfig{file=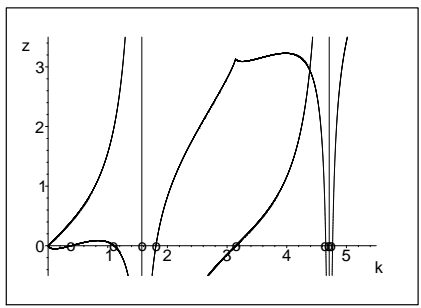,angle=270,width=0.4\textwidth}
\end{center}                         %instead of \end{center}
\vspace{-2mm} \caption{The real-momentum-dependence of the two
factors of the secular determinant of Eq.~(\ref{bgein6s})
 in the subcritical non-Hermitian regime with
$\alpha=0.7$.
 \label{fixfdubr}}
\end{figure}

\begin{figure}[h]                     %instead of \begin{figure}[t]
\begin{center}                         %instead of \begin{center}
\epsfig{file=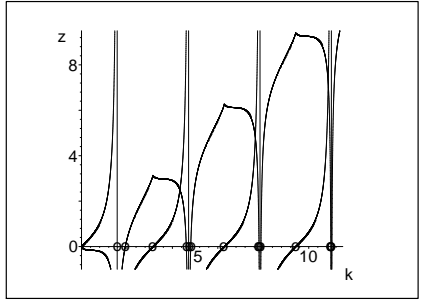,angle=270,width=0.4\textwidth}
\end{center}                         %instead of \end{center}
\vspace{-2mm} \caption{The real-momentum-dependence of the two
factors of the secular determinant of Eq.~(\ref{bgein6s})
 in the strongly non-Hermitian regime with
$\alpha=1$.
 \label{fixfdu}}
\end{figure}

The main technical {\em disadvantage} of the discretization recipe
lies in the quick weakening of the merits (viz., of the tractability
of the mathematics by means of linear algebra) with the growth of
$q$ and/or of the matrix dimensions (say, beyond  $N > 10$
\cite{fundgra}). The occurrence of these limitations forced us to
return, recently, to the continuous graphs of the star-shaped form
which is sampled in Fig.~\ref{fixu}. In Ref.~\cite{archiv}, in
particular, we omitted the ``central-vertex'' interaction (as used
in Ref.~\cite{fundgra}) completely. This means that just the most
elementary Kirchhof's law has been postulated to hold in the $x=L$
center of the graph with the wedges oriented, conventionally,
inwards,
 \begin{equation}
 \label{bcinge}
  \psi_{j}(L)=\psi_{0}(L)\,,\ \ \ \ j =  1, \ldots, q-1\,,
 \ \ \ \ \ \ \
  \sum_{j=0}^{q-1}\,\partial_{x_{j}}\psi_{j}(L)=0
  \,.
 \end{equation}
The simplification has been compensated by the
``exotic''\footnote{An anonymous referee of this paper correctly
objected that these boundary conditions need not be considered
exotic at all. The essence of such an objection may be found
explained in paper \cite{referee} in which the authors show how, in
the $q=2$ special case, these boundary conditions may be given a
natural and elementary physical interpretation in terms of a
perfect-transmission external scattering. Naturally, the presence of
the additional phase factors still leaves the net inflow into the
whole graph equal to zero at any $q>2$. Another extremely
interesting aspect of the models of similar class has also been
pointed out and discussed (though not yet published) by Stefan
Rotter et al \cite{Rotter}} external boundary conditions
 \begin{equation}
 \label{brody}
  \partial_{x_{j}}\psi_{j}(0)= i \alpha\, e^{{\rm i}j\varphi}\,
  \psi_{j}(0)\,,
  \ \ \ \ \ j = 0, 1, \ldots, q-1\,,
  \ \ \ \varphi=2\pi/q
  \,.
 \end{equation}
In the units $L=1$ we arrived at the $q=6$ secular equation
 \be
  \label{bgein6s}
  \frac{
 k^6-\alpha^6\, \tan^{4} k}{k^6+\alpha^6 \tan^6 k
    }\,\tan k
    =0\,.
 \ee
This equation admits the straightforward graphical solution which is
being sampled here via Figs.~\ref{fixfdubr} and ~\ref{fixfdu}.

%loweral.txt
%\newpage
%********** Figure 0 zde
%\newpage
%m:=2;alp:=0.7;el:=1; plot({(k -alp
% *(abs(tan(k*el)))^((4*m-4)/(4*m-2))),tan(k*el)},k=0..5.5,z=-0.5..3.5,numpoints=10000,
% scaling=CONSTRAINED    );

%loweral.txt
%\newpage
%********** Figure 0 zde
%\newpage
%m:=2;alp:=0.7;el:=1; plot({(k -alp
% *(abs(tan(k*el)))^((4*m-4)/(4*m-2))),tan(k*el)},k=0..5.5,z=-0.5..3.5,numpoints=10000,
% scaling=CONSTRAINED    );

A thorough inspection of these two pictures reveals that there
exists the so called Kato's \cite{Kato} exceptional-point value of
the critical coupling which may be localized to lie at
$\alpha_{min}\approx 0.7862806298$. At this value of the coupling
(which is, in our model, real) the number of the real roots of the
secular determinant changes. More precisely, the more detailed
Fig.~\ref{fixfdubr} shows that at the not too large value of
coupling $\alpha = 7/10$ there exist four distinct real zeros of the
secular determinant in the interval of $k \in (0,2)$. The subsequent
inspection of Fig.~\ref{fixfdu} enables us to find out that after
the transition to an overcritical value of $\alpha = 1$ the two
lowest roots disappeared and, obviously, complexified (in an
independent calculation, the position of these ``missing roots'' in
the complex plane of $k$ has been localized by an {\em ad hoc},
brute-force numerical method).

\section{Summary and outlook}

The idea of the study of the non-Hermitian star-shaped quantum
graphs relates, equally strongly, to the studies of its
non-Hermitian predecessors which remain topologically trivial (cf.
their samples, say, in Refs. \cite{Carl,SIGMA} and \cite{jakub})
{\em and} to the studies of the topologically nontrivial, genuine
$q-$pointed-star-shaped quantum graphs with $q>2$ where the
Hamiltonian $H$ of the quantum system is assumed self-adjoint in one
of the most common Hilbert-space forms of square-integrable wave
functions (see, {\em pars pro toto}, an overall discussion of this
approach in recent paper \cite{Nizhnik} which lists also further
references in this area). Each of these predecessors offers a
slightly different motivation which originated much more from the
needs of the phenomenology and quantum theory in the former case
(cf., in particular, \cite{Carl}) and which was able to make use of
an extensive knowledge of the underlying mathematics and, in
particular, of the existing theorems and results of functional
analysis in the latter context.

In this sense, at present as well as in the nearest future, the
combination of the two ideas may be expected to share both their
merits and weaknesses. In particular, the strengths may be expected
to emerge due to the related new perspectives in the applications in
physics (cf. e.g., the new perspective which appeared opening in the
experimental optics using newly developed anomalous metamaterials
\cite{Makris}). At the same time, the most important weak points of
the theory may, understandably, be identified with the current lack
of the reliable abstract mathematical understanding of some
subtleties which naturally emerge not only in connection with the
possible, uninhibited encounter of the Kato's exceptional points,
say, in the space of couplings, but also in connection with our --
up to now fairly vague -- current understanding of the
representation theory of Hilbert spaces in which the operator of
metric is admitted non-trivial (cf., e.g., \cite{Siegl} for further
reading).

\vspace{5mm}

%\newpage

\subsection*{Acknowledgements}

The support by the GA\v{C}R grant Nr. P203/11/1433 is acknowledged.
The author also appreciates discussions with Ondrej Turek and Taksu
Cheon (Kochi University), with Sergii Kuzhel (AGH University in
Cracow) and, last but not least, unidirectionally, with anonymous
referees.


\newpage

\begin{thebibliography}{00}


\bibitem{Exner}
P. Exner, J. P. Keating, P. Kuchment and A. Teplyaev, Analysis on
Graphs and Its Applications (AMS, Rhode Island, 2008).

\bibitem{Carl}
C. M. Bender, Rep. Prog. Phys. 70 (2007) 947.

\bibitem{SIGMA}
M. Znojil,
%Three-Hilbert-space formulation of Quantum Mechanics
%SYMMETRY, INTEGRABILITY and GEOMETRY: METHODS and APPLICATIONS
SIGMA 5,  001 (2009) (arXiv overlay: 0901.0700).
%, 19 pages ();%
%
%Miloslav Znojil, Three-Hilbert-space formulation of Quantum
%Mechanics SYMMETRY, INTEGRABILITY and GEOMETRY: METHODS and
%APPLICATIONS SIGMA 5 (2009), 001, 19 pages; (doi:
%10.3842/SIGMA.2009.001) Contribution to the Proceedings issue
%dedicated to the VIIth Workshop ''Quantum Physics with Non-Hermitian
%Operators'' (arXiv:0901.0700 [quant-ph, math-ph, math.MP] 6 Jan
%2009)


\bibitem{jakub}
J. \v{Z}elezn\'y, The Krein-space theory for non-Hermitian
PT-symmetric operators (MSc thesis, FNSPE CTU, 2011);

P. Siegl,
 Non-Hermitian quantum models, indecomposable
representations and coherent states quantization (PhD thesis, Univ.
Paris Diderot  \& FNSPE CTU, 2011).




\bibitem{Smilga}
M. Znojil,
%Miloslav Znojil, Time-Dependent and/or Nonlocal Representations of
%Hilbert Spaces in Quantum Theory
Acta Polytechnica 50 (2010) 62.

\bibitem{Dieudonne}
 J. Dieudonne,  %Quasi-Hermitian operators,
 Proc. Int. Symp.
            Lin. Spaces (Pergamon, Oxford,   1961), p. 115.
            %-122

\bibitem{Geyer}
F. G. Scholtz, H. B. Geyer  and F. J. W. Hahne, Ann. Phys. (NY) 213
 (1992) 74.
%% Quasi-Hermitian Operators in Quantum Mechanics
%% and the Variational Principle,

\bibitem{DDT}
P. Dorey, C. Dunning and R. Tateo,
% [ps, pdf, other] Title: The ODE/IM
%Correspondence Authors: Patrick Dorey, Clare Dunning, Roberto Tateo
%Comments: 102 pages, 27 figures. v2: added references Journal-ref:
%
J. Phys. A: Math. Theor. 40 (2007) R205
%(arXiv:hep-th/0703066)
and
% [ps, pdf, other]
%Title: From PT-symmetric quantum mechanics to conformal field theory
%Authors:
Pramana - J. Phys. 73 (2009) 217.
%, to appear %Comments: 27
%pages, 12 figures, contribution to the Proceedings of PHHQP VIII
%(Mumbai, January 2009)
% (arXiv:0906.1130).

\bibitem{cubic}
A. Mostafazadeh, Int. J. Geom. Meth. Mod. Phys. 7 (2010) 1191.



\bibitem{fund}
M. Znojil, %Fundamental length in quantum theories with PT-symmetric
%Hamiltonians
Phys. Rev. D. 80 (2009) 045022.
% (13
%pages) http://dx.doi.org/10.1103/PhysRevD.80.045022 (arXiv:0907.2677
%[hep-th] 15 Jul 2009)


\bibitem{fundgra}
M. Znojil, %Fundamental length in quantum theories with PT-symmetric
%Hamiltonians II: The case of quantum graphs.
Phys. Rev. D. 80 (2009) 105004.
%105004 ( No. 10, 15 November 2009, 20 pages)
%http://dx.doi.org/10.1103/PhysRevD.80.105004 (arXiv:0910.2560
%[hep-th] 14 Oct 2009)


\bibitem{katka}
P. Exner and K. N\v{e}mcov\'{a}, J. Phys. A: Math. Gen. 36 (2003)
10173.

\bibitem{anomalous}
M. Znojil,
%
%Anomalous real spectra of non-Hermitian quantum graphs in
%strong-coupling regime Authors: Miloslav Znojil Comments: 17 pp., 9
%figures, published version Journal-ref:
J. Phys. A: Math. Theor. 43 (2010) 335303;

M. Znojil,
%Cryptohermitian Hamiltonians on graphs
Int. J. Theor. Phys. 50 (2011) 1052 and 1614.

\bibitem{archiv}
M. Znojil,
%PT-symmetry and quantum graphs
%(arXiv:1205.5211).
%Miloslav Znojil,
%"Quantum star-graph analogues of PT-symmetric square wells".
Can. J. Phys. 90 (2012) 1287.
%-1293 Editor s choice: open access:
%http://dx.doi.org/10.1139/p2012-107 (arXiv:1205.5211)

\bibitem{smeared}
M. Znojil,
%Scattering theory using smeared non-Hermitian potentials
Phys. Rev. D. 80 (2009) 045009.


\bibitem{kappa}
M. Znojil, %On the role of the normalization factors $\kappa_n$ and
%of the pseudo-metric P in crypto-Hermitian quantum models. SYMMETRY,
%INTEGRABILITY and GEOMETRY: METHODS and APPLICATIONS
SIGMA 4, 001 (2008)
%001, 9 pages; (
(arXiv overlay: 0710.4432v3).
% [math-ph] 2 Jan 2008)

\bibitem{referee}
%
%``Perfect transmission scattering as a PT-symmetric spectral problem
H. Hernandez-Coronado, D. Krejcirik and P. Siegl, Phys. Lett. A 375
(2011) 2149.
%
%-2152 arXiv:1011.4281, animation

\bibitem{Rotter}
S. Rotter, ``Exceptional points in open and closed gain-loss
structures'',
http://phhqp11.in2p3.fr/Wednesday\_29\_files/RotterSL12.pdf

\bibitem{Kato}
T. Kato, Perturbation theory for linear operators (Springer, Berlin,
1966).

\bibitem{Nizhnik}
L. P.  Nizhnik, %L. P. Inverse eigenvalue problems for nonlocal
%Sturm-Liouville operators on a star graph.
Methods Funct. Anal. Topology 18 (2012) 68.
%, no. 1, 68–78.

\bibitem{Makris}
C. E. R\"{u}ter, R. Makris, K. G. El-Ganainy, D. N. Christodoulides,
M. Segev
and D. Kip,
% D 2010 %Observation of parity-time symmetry in optics
Nat. Phys. 6 (2010) 192.

\bibitem{Siegl}
F. Bagarello and M. Znojil, %"Non linear pseudo-bosons versus hidden
%Hermiticity. II: The case of unbounded operators"
J. Phys. A: Math. Theor. 45 (2012) 115311;
% doi:
%http://dx.doi.org/10.1088/1751-8113/45/11/115311 arXiv: 1202.2028

P. Siegl and D. Krejcirik,
%
%: "On the metric operator for the
%imaginary cubic oscillator";
Phys. Rev. D 86 (2012) 121702(R).
%Preprint available on arXiv:1208.1866 [quant-ph].

\end{thebibliography}
\end{document}